\documentclass[acmsmall]{acmart}
\AtBeginDocument{%
  \providecommand\BibTeX{{%
    \normalfont B\kern-0.5em{\scshape i\kern-0.25em b}\kern-0.8em\TeX}}}

\usepackage{hhline}
\usepackage{hyperref}

\usepackage{multirow}
\usepackage[table,xcdraw]{xcolor}
\usepackage{caption} 
\usepackage{subcaption}

\setcopyright{rightsretained}
\acmYear{2025} \acmVolume{1} \acmNumber{1} \acmArticle{1} \acmMonth{1} 
\acmPrice{}
\acmDOI{10.1145/3720536}

\begin{document}

\title{Generative AI Policies under the Microscope: How CS Conferences Are Navigating the New Frontier in Scholarly Writing}

\author{Mahjabin Nahar}
\email{mahjabin.n@psu.edu}
\affiliation{%
  \institution{The Pennsylvania State University}
  \city{University Park}
  \state{PA}
  \country{USA}
}

\author{Sian Lee}
\affiliation{%
  \institution{University of Mississippi}
  \city{Oxford}
  \state{MS}
  \country{USA}
}

\author{Rebekah Guillen}
\email{rdguil777@gmail.com}
\affiliation{%
  \institution{The MITRE Corporation}
  \city{San Antonio}
  \state{TX}
  \country{USA}
}

\author{Dongwon Lee}
\affiliation{%
  \institution{The Pennsylvania State University}
  \city{University Park}
  \state{PA}
  \country{USA}
}


\begin{abstract}
\textbf{Abstract.} As the use of Generative AI (Gen-AI) in scholarly writing and peer reviews continues to rise, it is essential for the computing field to establish and adopt clear Gen-AI policies. This study examines the landscape of Gen-AI policies across 64 major Computer Science conferences and offers recommendations for promoting more effective and responsible use of Gen-AI in the field.
\end{abstract}
\maketitle

\section{Landscape of Generative AI Policies in CS Conferences}

\begin{table}[t]

\caption{Leniency ratings on a 5-pt Likert scale, `1'=`Extremely restrictive' (restricts almost all types of use), `2'=`Somewhat restrictive' (restricts most types of use, but allows a few, e.g., allows grammar/spelling edits), `3'=`Neither lenient nor restrictive' (allows some types of use, restricts others, e.g., editing or polishing author-written content, may imply rewrite such as changing writing style, summarizing, etc.), `4'=`Somewhat lenient' (allows most types of use with detailed reporting, i.e., in text, figure, code, etc. with disclosures), `5'=`Extremely lenient' (almost all types of use, no reporting). Conferences that adopted society-level policies for authors are marked with an asterisk (*) and those with no policies are marked as `-'. Conference policies were initially evaluated between September 6-12, 2024, and subsequently reviewed and revised between December 12-17, 2024.}
\scalebox{0.68}{

\begin{tabular}{llccccc}
\hhline{-------}
\rowcolor[HTML]{FFFFFF} 
\cellcolor[HTML]{FFFFFF} & \cellcolor[HTML]{FFFFFF} & \cellcolor[HTML]{FFFFFF} & \multicolumn{2}{c}{\cellcolor[HTML]{FFFFFF}Year 1 (2023/2024)} & \multicolumn{2}{c}{\cellcolor[HTML]{FFFFFF}Year 2 (2024/2025)} \\ \hhline{~~~----}
\rowcolor[HTML]{FFFFFF} 
\multirow{-2}{*}{\cellcolor[HTML]{FFFFFF}Areas} & \multirow{-2}{*}{\cellcolor[HTML]{FFFFFF}Conference/Society Name/(s)} & \multirow{-2}{*}{\cellcolor[HTML]{FFFFFF}Year/(s)} & Author & Reviewer & Author & Reviewer \\ \hhline{-------}
\cellcolor[HTML]{FFFFFF} & \cellcolor[HTML]{FFFFFF}AAAI & \cellcolor[HTML]{FFFFFF}2024 & \multicolumn{2}{c}{\cellcolor[HTML]{D5D5D5}-} & \cellcolor[HTML]{CDE683}4 & \cellcolor[HTML]{FFCE93}2 \\
\cellcolor[HTML]{FFFFFF} & \cellcolor[HTML]{FFFFFF}ACM & \cellcolor[HTML]{FFFFFF}2024 & \multicolumn{2}{c}{\cellcolor[HTML]{D5D5D5}-} & \cellcolor[HTML]{CDE683}4 & \cellcolor[HTML]{FFFC9E}3 \\
\rowcolor[HTML]{FFFFFF} 
\multirow{-3}{*}{\cellcolor[HTML]{FFFFFF}\begin{tabular}[c]{@{}l@{}}Society (3)\end{tabular}} & IEEE & 2024 & \multicolumn{2}{c}{\cellcolor[HTML]{D5D5D5}-} & \cellcolor[HTML]{CDE683}4 & - \\ \hhline{-------}
\rowcolor[HTML]{FFFFFF} 
\cellcolor[HTML]{FFFFFF} & AAAI & 2024, 2025 & \cellcolor[HTML]{FFFC9E}3 & - & \cellcolor[HTML]{FFFC9E}3 & - \\
\rowcolor[HTML]{FFFFFF} 
\cellcolor[HTML]{FFFFFF} & IJCAI & 2024, 2025 & \cellcolor[HTML]{FFFC9E}3 & - & \cellcolor[HTML]{FFFC9E}3 & - \\
\rowcolor[HTML]{FFFFFF} 
\cellcolor[HTML]{FFFFFF} & CVPR & 2024, 2025 & \cellcolor[HTML]{92FE8F}5 & \cellcolor[HTML]{FFFC9E}3 & \cellcolor[HTML]{92FE8F}5 & \cellcolor[HTML]{FFCE93}2 \\
\cellcolor[HTML]{FFFFFF} & \cellcolor[HTML]{FFFFFF}ECCV & \cellcolor[HTML]{FFFFFF}2024 & \multicolumn{2}{c}{\cellcolor[HTML]{D5D5D5}-} & \cellcolor[HTML]{92FE8F}5 & \cellcolor[HTML]{FFFC9E}3 \\
\rowcolor[HTML]{FFFFFF} 
\cellcolor[HTML]{FFFFFF} & ICCV & 2023 & \multicolumn{2}{c}{\cellcolor[HTML]{D5D5D5}-} & - & \cellcolor[HTML]{FA9A94}{\color[HTML]{000000} 1} \\
\rowcolor[HTML]{FFFFFF} 
\cellcolor[HTML]{FFFFFF} & ICLR & 2024, 2025 & \cellcolor[HTML]{FFFC9E}3 & - & \cellcolor[HTML]{FFFC9E}3 & \cellcolor[HTML]{FFFC9E}3 \\
\rowcolor[HTML]{FFFFFF} 
\cellcolor[HTML]{FFFFFF} & ICML & 2024, 2025 & \cellcolor[HTML]{FFFC9E}3 & - & \cellcolor[HTML]{FFFC9E}3 & - \\
\cellcolor[HTML]{FFFFFF} & \cellcolor[HTML]{FFFFFF}NeurIPS & \cellcolor[HTML]{FFFFFF}2023, 2024 & \cellcolor[HTML]{CDE683}4 & - & \cellcolor[HTML]{CDE683}4 & \cellcolor[HTML]{FFFFFF}- \\
\rowcolor[HTML]{FFFFFF} 
\cellcolor[HTML]{FFFFFF} & ACL & 2024, 2025 & \cellcolor[HTML]{CDE683}4 & - & \cellcolor[HTML]{CDE683}4 & - \\
\rowcolor[HTML]{FFFFFF} 
\cellcolor[HTML]{FFFFFF} & EMNLP & 2023, 2024 & - & - & \cellcolor[HTML]{CDE683}4 & - \\
\rowcolor[HTML]{FFFFFF} 
\cellcolor[HTML]{FFFFFF} & WWW & 2024, 2025 & \cellcolor[HTML]{CDE683}4 & - & \cellcolor[HTML]{FFFC9E}3 & - \\
\rowcolor[HTML]{FFFFFF} 
\cellcolor[HTML]{FFFFFF} & NAACL & 2025 & - & - & \cellcolor[HTML]{CDE683}4 & - \\
\rowcolor[HTML]{FFFFFF} 
\multirow{-12}{*}{\cellcolor[HTML]{FFFFFF}AI (13)} & SIGIR* & 2024, 2025 & - & - & \cellcolor[HTML]{CDE683}4 & - \\ 
\hhline{-------}
\rowcolor[HTML]{FFFFFF} 
\cellcolor[HTML]{FFFFFF} & SIGCSE* & 2024, 2025 & - & - & \cellcolor[HTML]{CDE683}4 & \cellcolor[HTML]{FFFC9E}3 \\
\cellcolor[HTML]{FFFFFF} & \cellcolor[HTML]{FFFFFF}CHI & \cellcolor[HTML]{FFFFFF}2024, 2025 & \cellcolor[HTML]{CDE683}4 & \cellcolor[HTML]{FFFC9E}3 & \cellcolor[HTML]{CDE683}4 & \cellcolor[HTML]{FFFC9E}3 \\
\rowcolor[HTML]{FFFFFF} 
\cellcolor[HTML]{FFFFFF} & UbiComp / Pervasive / IMWUT* & 2023, 2024 & - & - & \cellcolor[HTML]{CDE683}4 & \cellcolor[HTML]{FFCE93}2 \\
\rowcolor[HTML]{FFFFFF} 
\cellcolor[HTML]{FFFFFF} & UIST & 2024, 2025 & \cellcolor[HTML]{FA9A94}1 & - & \cellcolor[HTML]{FA9A94}1 & - \\
\rowcolor[HTML]{FFFFFF} 
\cellcolor[HTML]{FFFFFF} & RSS & 2024, 2025 & \cellcolor[HTML]{FFFC9E}3 & - & \cellcolor[HTML]{FFFC9E}3 & - \\
\rowcolor[HTML]{FFFFFF} 
\cellcolor[HTML]{FFFFFF} & VIS* & 2023, 2024 & - & - & \cellcolor[HTML]{CDE683}4 & \cellcolor[HTML]{FFCE93}2 \\
\rowcolor[HTML]{FFFFFF} 
\cellcolor[HTML]{FFFFFF} & VR & 2024, 2025 & \cellcolor[HTML]{FFCE93}2 & - & \cellcolor[HTML]{FFCE93}2 & - \\
\rowcolor[HTML]{FFFFFF} 
\cellcolor[HTML]{FFFFFF} & ICRA & 2024, 2025 & - & - & \cellcolor[HTML]{CDE683}4 & \cellcolor[HTML]{FA9A94}1 \\
\rowcolor[HTML]{FFFFFF} 
\cellcolor[HTML]{FFFFFF} & IROS & 2024, 2025 & - & - & \cellcolor[HTML]{CDE683}4 & \cellcolor[HTML]{FA9A94}1 \\
\multirow{-8}{*}{\begin{tabular}[c]{@{}l@{}}Interdis-\\ ciplinary (15)\end{tabular}} & \begin{tabular}[c]{@{}l@{}}ISMB, RECOMB, SIGGRAPH,\\ SIGGRAPH Asia, EC, WINE\end{tabular} & \begin{tabular}[c]{@{}l@{}}2023, 2024,\\2025\end{tabular} & - & - & - & - \\
\hhline{-------}
\rowcolor[HTML]{FFFFFF} 
\cellcolor[HTML]{FFFFFF} & SIGCOMM* & 2024, 2025 & \cellcolor[HTML]{CDE683}4 & - & \cellcolor[HTML]{CDE683}4 & - \\
\rowcolor[HTML]{FFFFFF} 
\cellcolor[HTML]{FFFFFF} & SIGMOD* & 2024, 2025 & - & - & \cellcolor[HTML]{CDE683}4 & - \\
\rowcolor[HTML]{FFFFFF} 
\cellcolor[HTML]{FFFFFF} & DAC* & 2024, 2025 & \cellcolor[HTML]{CDE683}4 & - & \cellcolor[HTML]{CDE683}4 & - \\
\rowcolor[HTML]{FFFFFF} 
\cellcolor[HTML]{FFFFFF} & HPDC & 2024, 2025 & \cellcolor[HTML]{CDE683}4 & - & \cellcolor[HTML]{FFFC9E}3 & - \\
\rowcolor[HTML]{FFFFFF} 
\cellcolor[HTML]{FFFFFF} & SC & 2023, 2024 & \cellcolor[HTML]{CDE683}4 & - & \cellcolor[HTML]{FFFC9E}3 & - \\
\rowcolor[HTML]{FFFFFF} 
\cellcolor[HTML]{FFFFFF} & IMC* & 2024, 2025 & \cellcolor[HTML]{CDE683}4 & \cellcolor[HTML]{FFFC9E}3 & \cellcolor[HTML]{CDE683}4 & \cellcolor[HTML]{FFFC9E}3 \\
\rowcolor[HTML]{FFFFFF} 
\cellcolor[HTML]{FFFFFF} & FSE* & 2024, 2025 & \cellcolor[HTML]{CDE683}4 & - & \cellcolor[HTML]{CDE683}4 & - \\
\rowcolor[HTML]{FFFFFF} 
\cellcolor[HTML]{FFFFFF} & ICSE* & 2024, 2025 & - & - & \cellcolor[HTML]{CDE683}4 & - \\
\rowcolor[HTML]{FFFFFF} 
\cellcolor[HTML]{FFFFFF} & NSDI & 2024, 2025 & - & - & \cellcolor[HTML]{FFFC9E}3 & - \\
\rowcolor[HTML]{FFFFFF} 
\cellcolor[HTML]{FFFFFF} & MobiSys* & 2024, 2025 & - & - & \cellcolor[HTML]{CDE683}4 & - \\
\multirow{-9}{*}{Systems (29)} & \begin{tabular}[c]{@{}l@{}}ASPLOS, ISCA, MICRO, CCS,\\IEEE S\&P ("Oakland"), USENIX Security,\\ VLDB, ICCAD, EMSOFT, RTAS, ICS,\\MobiCom, SenSys, SIGMETRICS, OSDI,\\SOSP, PLDI, POPL, RTSS\end{tabular} & \begin{tabular}[c]{@{}l@{}}2023, 2024,\\2025\end{tabular} & - & - & - & - \\ 
\hhline{-------}
Theory (7) & \begin{tabular}[c]{@{}l@{}}FOCS, SODA, STOC, \\ CRYPTO, EuroCrypt, \\ CAV, LICS\end{tabular} & \begin{tabular}[c]{@{}l@{}}2023, 2024,\\2025\end{tabular} & - & - & - & - \\ 
\hhline{-------} 

\end{tabular}
}
\label{tab:table}
\end{table}

Since the rise of ChatGPT, {\em Generative AI} (Gen-AI) technologies gained widespread popularity, impacting academic research and everyday communication~\citep{lin2024towards, uchendu2023attribution}. While Gen-AI offers benefits in task automation~\citep{HarvardBusinessReview},
it can also be misused and abused in nefarious applications~\cite{lee-cacm24}, with significant risks to long-tail populations~\cite{lee-comp}.
Professionals in fields like journalism and law still remain cautious due to concerns over hallucinations and ethical issues, 
but scholars in Computer Science (CS), the field where Gen-AI originated, appear to be cautiously, yet actively exploring its use. For instance, ~\cite{zou-colm} reports the increased use of large language models (LLMs) in the CS scholarly articles (up to 17.5\%), compared to Mathematics articles (up to 6.3\%), and \cite{liang2024monitoring} reports that,
between 6.5\% and 16.9\% of peer reviews at ICLR 2024, NeurIPS 2023, CoRL 2023, and EMNLP 2023 may have been altered by LLMs beyond minor revisions.

Considering researchers’ increasing adoption of Gen-AI, it is crucial to establish usage policies to promote fair and ethical practices in scholarly writing and peer reviews. Some conferences, such as \href{https://icml.cc/Conferences/2023/llm-policy}{ICML 2023} accurately highlighted the confusion and questions surrounding Gen-AI use, with concerns regarding the novelty and ownership of generated content. Previous research examined Gen-AI policies by major publishers like Elsevier, Springer, etc.~\citep{lin2024towards}, but there is still a lack of clear understanding of how CS conferences are adapting to this paradigm shift. Hence, this article provides a summarized overview of the ``scholarly writing'' policies of CS conferences and major computing societies: ACM, IEEE, and AAAI (2024 and 2025 if available; otherwise 2023 and 2024) and offers recommendations. Conferences studied in this analysis were selected based on \textit{csrankings}\footnote{https://csrankings.org/}, including leading conferences in each CS subfield. Our analysis shows that many CS conferences have not established Gen-AI policies and those that have, vary in leniency, disclosure, and sanctions. Policies for authors are more prevalent compared to reviewers, and some address code writing and documentation. These policies are evolving, as demonstrated by conferences such as ICML 2023, which initially prohibited LLM-generated text but later clarified its allowance for editing author-written content. However, by and large, adoption remains inconsistent across conferences, creating uncertainty in their application.

Table 1 lists the conferences considered, alongside their leniency ratings on a 5-point Likert scale. Three authors independently annotated the policies after agreeing on rating criteria. Given the ordinal nature of the ratings, we used Krippendorff’s alpha ($\alpha =0.832$, satisfactory) to assess inter-rater reliability. Final ratings were assigned through majority voting. Nevertheless, we acknowledge that variations in ratings may arise from the subjective nature of the task, largely attributable to the ambiguity in policy language.

Most conferences with Gen-AI policies were somewhat lenient for authors (ratings of `3' or `4'). The 2025 author policy for \href{https://www2025.thewebconf.org/research-tracks}{WWW} was assigned a rating of `3' as it allowed rephrasing and \href{https://2025.ieee-icra.org/announcements/paper-submissions/}{ICRA} was assigned `4' as it allowed different types of usage with disclosure. Interdisciplinary conferences like \href{https://uist.acm.org/2024/cfp/}{UIST} and \href{https://ieeevr.org/2025/contribute/papers/}{VR} had lower leniency ratings of `1' (`no AI-generated content allowed) and `2' (e.g., allowing grammar/spelling edits), respectively. In contrast, several AI conferences, including \href{https://cvpr.thecvf.com/Conferences/2025/AuthorGuidelines}{CVPR}, were highly permissive of Gen-AI use with rating `5' (e.g., no restrictions), likely due to a greater familiarity of the AI field with LLMs and their perceived benefits. In comparison, conferences adopted more restrictive policies for reviewers. Particularly, \href{https://2025.ieee-icra.org/announcements/paper-submissions/}{ICRA}, \href{https://ieeevis.org/year/2024/info/call-participation/review-instructions}{VIS}, and \href{https://conferences.sigcomm.org/imc/2025/submission-instructions/}{IMC} were rated `1', `2', and `3', respectively, likely stemming from concerns over content leakage and limitations of LLMs in handling complex tasks~\cite{wenpeng24}.

\section{Area-Level Trends}

\begin{figure}[h]
    \centering
    \begin{subfigure}{0.497\textwidth}
        \centering
        \includegraphics[width=\textwidth]{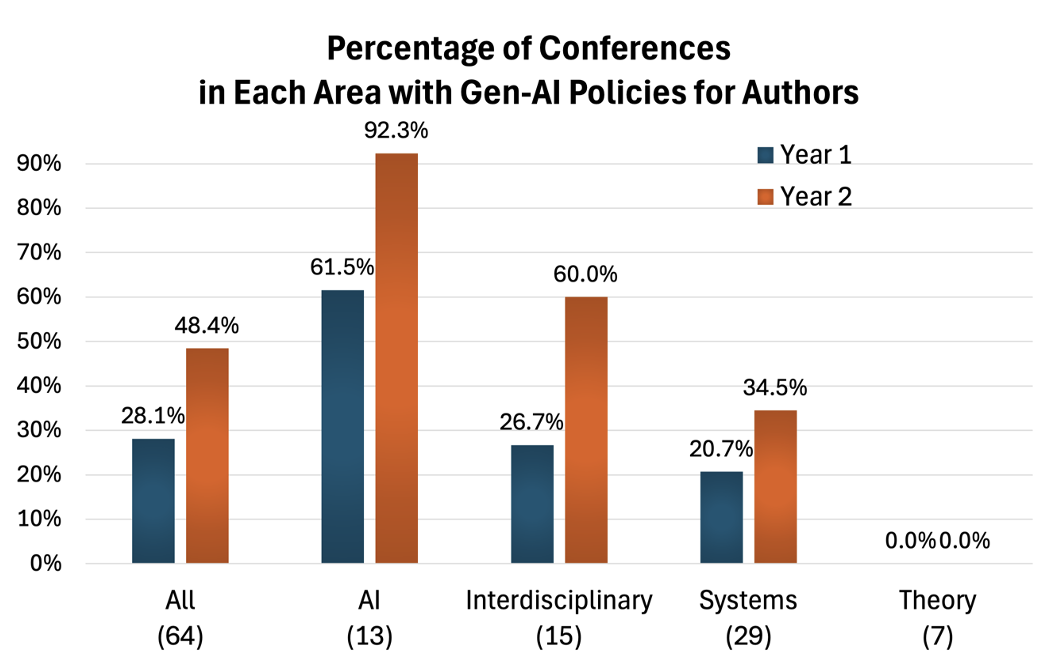} 
        \caption{For Authors}
        \label{fig:aut1}
    \end{subfigure}
    \hfill
    \begin{subfigure}{0.497\textwidth}
        \centering
        \includegraphics[width=\textwidth]{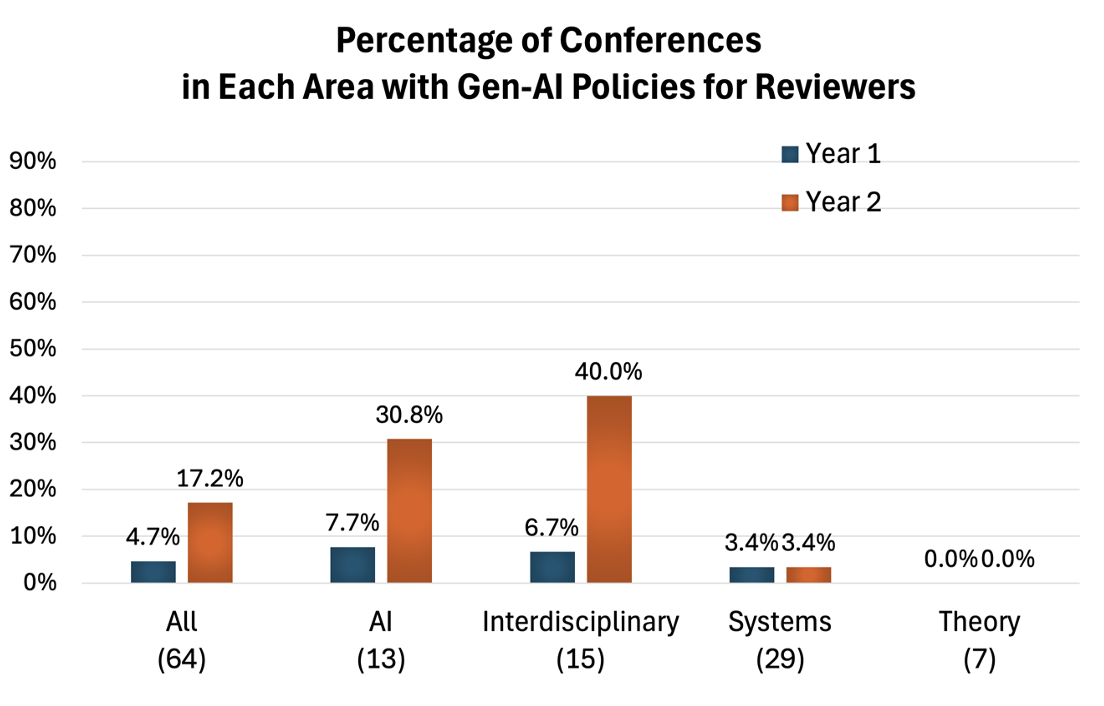}
        \caption{For Reviewers}
        \label{fig:rev1}
    \end{subfigure}
    \caption{Area-wise percentage (\%) of conferences with Gen-AI policies for authors and reviewers in Years 1 and 2. The number in parentheses indicates the number of conferences per area.}
    \label{fig:adoption}
\end{figure}

We compared the trend of Gen-AI policy adoption across different CS areas, following the classification from csrankings: \emph{AI}, \emph{Interdisciplinary}, \emph{Systems}, and \emph{Theory}. For {AI}, 61.5\% of conferences had Gen-AI policies for authors in Year 1, which increased to 92.3\% in Year 2 (Figure ~\ref{fig:aut1}), indicating that conferences in {AI} field are the most active in adopting Gen-AI policies for authors. Conferences in the Interdisciplinary area showed 26.7\% in Year 1 and 60\% in Year 2, making them the second most proactive in introducing Gen-AI policies for authors. While these two areas are leading in Gen-AI policy adoption for authors, the {Systems} area lagged behind the overall trend in Year 1 (20.7\%) but saw a 13.8\% increase in Year 2 (34.5\%), implying that this area is gradually aligning with the overall trend. A similar pattern was observed for reviewers, with AI (Year 1: 7.7\%, Year 2: 30.8\%) and Interdisciplinary (Year 1: 6.7\%, Year 2: 40\%) leading Gen-AI policy adoption, while the Systems area (Year 1: 3.4\%, Year 2: 3.4\%) is gradually converging with the broader trend. In contrast, no conferences in the {Theory} area had Gen-AI policies for authors or reviewers. This may be due to the conservative nature of the area or lack of active Gen-AI usage in writing theory articles.

\begin{figure}[h]
    \centering
    \begin{subfigure}{0.48\textwidth}
        \centering
        \includegraphics[width=\textwidth]{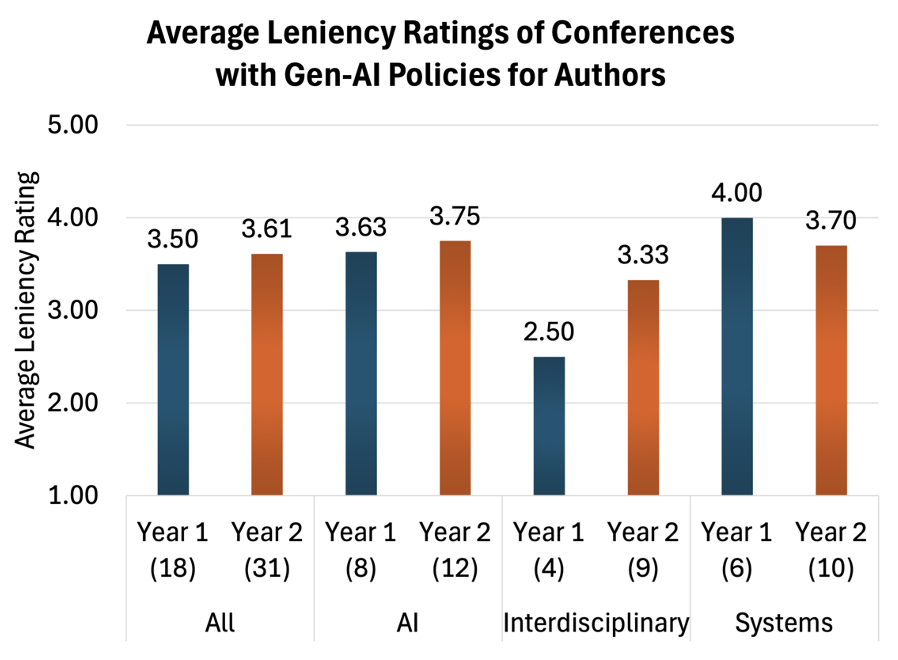} 
        \caption{For Authors}
        \label{fig:aut2}
    \end{subfigure}
    \hfill
    \begin{subfigure}{0.48\textwidth}
        \centering
        \includegraphics[width=\textwidth]{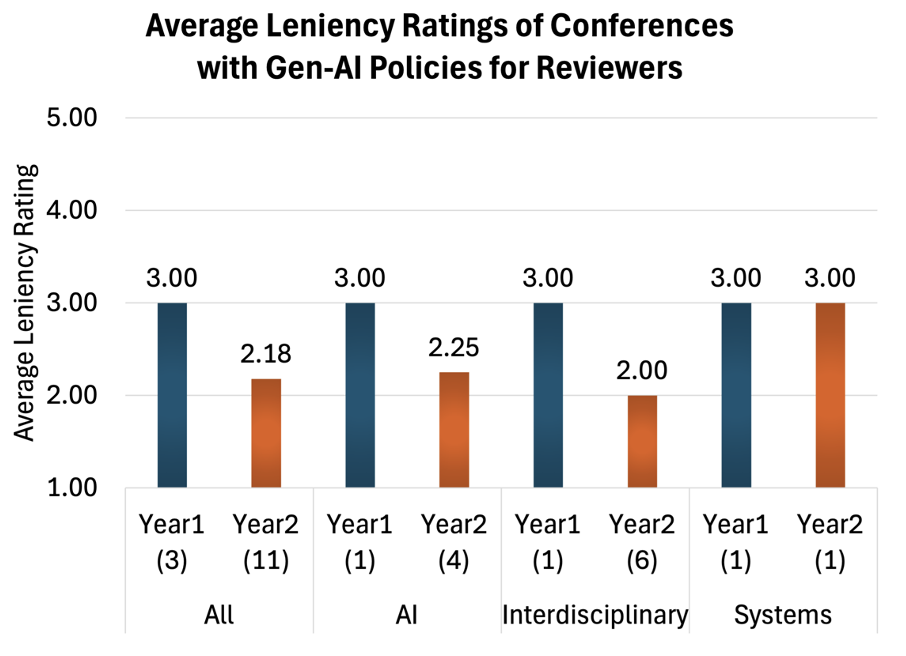}
        \caption{For Reviewers}
        \label{fig:rev2}
    \end{subfigure}
    \caption{Area-wise average leniency ratings of conferences with Gen-AI policies for authors and reviewers in Years 1 and 2. The ratings were calculated based only on conferences that adopted Gen-AI policies for each year in each area. Theory field was excluded as those conferences did not adopt Gen-AI policies.}
    \label{fig:leniency}
\end{figure}

In terms of author policies, {AI} area shows higher leniency ratings compared to the overall average (see Figure~\ref{fig:aut2}). Additionally, the {Systems} area shows very high leniency ratings (Year 1: 4.00, Year 2: 3.7) compared to overall average (Year 1: 3.5, Year 2: 3.61). In contrast, the leniency ratings in the {Interdisciplinary} area are relatively lower (Year 1: 2.50, Year 2: 3.33). This may imply that, despite being more proactive in adopting policies than Systems, Interdisciplinary area tends to be more cautious, imposing stricter guidelines on the use of Gen-AI.

While conferences with Gen-AI policies tend to be more lenient toward authors, with an average leniency rating of 3.50 in Year 1 and 3.61 in Year 2 - both above 3 - the policies for reviewers are notably more restrictive (see Figure~\ref{fig:rev2}). The average leniency rating for reviewers dropped from 3.00 in Year 1 to 2.18 in Year 2, falling below 3, and no conferences granted reviewers a leniency rating above 3 (see Table~\ref{tab:table}). This likely reflects concerns about the potential risks of using AI tools in the peer review process, such as the possibility of leaking sensitive or unpublished information and inadequate capability of LLMs  for performing tasks of a high level of expertise and nuanced judgment~\cite{wenpeng24}. Since reviewers have access to confidential work, conferences may be more cautious to ensure data security, privacy, and the protection of intellectual property.

\section{Temporal Trends}
We examined the Gen-AI policies for two consecutive years, 2024 and 2025, and when unavailable, 2023 and 2024. Overall, conferences are moving toward introducing new Gen-AI policies. In Year 1, only 18 (28.1\%) out of 64 conferences had Gen-AI policies, either for authors or reviewers, which increased to 32 (50\%) in Year 2. Specifically, for authors (see Figure~\ref{fig:aut1}), only 18 (28.1\%) had Gen-AI policies in Year 1, increasing to 31 (48.4\%) in Year 2, a 20.3 percentage point increase. In contrast, for reviewers (see Figure~\ref{fig:rev1}), only 3 (4.7\%) conferences had Gen-AI policies in Year 1, rising to 11 (17.2\%) in Year 2. While many conferences are increasingly adopting and publicly sharing Gen-AI policies for authors, they are slower in providing clear guidelines for reviewers. Conferences may be unaware of reviewers' needs or could be implementing Gen-AI policies for reviewers internally without making them publicly available. In either case, there is a noticeable disparity in how Gen-AI policies are adopted and communicated for authors versus reviewers.

Several conferences adopted Gen-AI policies during the latter year, including SIGMOD and EMNLP for authors, and SIGCSE and VIS for authors and reviewers. Once a policy was established, similar degree of leniency was maintained for the latter year, with a few exceptions (e.g., SC shifted to a more restrictive policy in 2024). These findings suggest a cautious, evolving approach to Gen-AI policies, with some conferences adopting clearer guidelines with increasing Gen-AI use.

\section{Society vs. Conference-Level Trends}
We reviewed the Gen-AI policies of ACM, IEEE, and AAAI, as many CS conferences are affiliated with these societies. While conference-specific policies differ in disclosure requirements and flexibility, society-level policies permit authors' Gen-AI use with disclosures. ACM and AAAI also permit reviewers to enhance their reviews using Gen-AI, as long as the submissions remain unexposed to these systems. In contrast, IEEE has no policies for conference reviewers. None of these policies mention sanctions for authors or reviewers. As Table~\ref{tab:table} summarizes, some conferences do not have conference-specific Gen-AI policies for authors and refer to the society-level policies--e.g., SIGCSE, VIS, IMC, while others have established their own Gen-AI policies for authors, including CVPR (IEEE) and UIST (ACM). Still, many conferences appear unaware of society-level Gen-AI policies, as the conference websites do not mention them. For reviewers, few conferences adhere to society-level policies (e.g., IMC), while many follow their own (e.g., CVPR).

\section{Recommendations}
Among the 64 conferences analyzed, 32 (50\%) adopted Gen-AI policies for authors or reviewers in two years, but only 11 (17.18\%) addressed reviewers. {Theory} and {Systems} fields lagged behind, with no {Theory} and 34.5\% {Systems} conferences adopting policies for authors by Year 2, compared to 92.3\% in {AI} and 60\% in {Interdisciplinary} fields. Notably, 14 of 32 policies emerged in Year 2, and no conference allowed Gen-AI authorship. Many ACM and IEEE conferences appear unaware of society-level policies, showing inconsistent adoption. Policies also lack sanctions for non-compliance, which is essential to enforce rules and prevent misuse.

A notable gap exists in reviewers' Gen-AI policies, perhaps stemming from concerns about exposing sensitive information. Yet, this risk can be mitigated by configuring LLM settings to ensure models neither retain nor learn from user-provided data \cite{lin2024beyond}. LLMs, e.g., OpenAI's ChatGPT and Google's Gemini, allow users to opt out of model training, providing added safeguards \cite{lin2024beyond}. Moreover, LLMs provide enterprise versions with data security (e.g. ChatGPT\footnote{https://openai.com/enterprise-privacy/} and Anthropic's Claude\footnote{https://privacy.anthropic.com/en/collections/10672411-data-handling-retention}) or accommodate privacy requests\footnote{https://privacy.openai.com/policies}. With the rise of Gen-AI, conferences without policies should establish guidelines for authors and reviewers, including evaluating AI use in submissions. In addition, Gen-AI policies at the area or society level would ensure consistency, benefiting both authors and reviewers, especially for resubmissions. Moreover, implementing professional development initiatives, training programs, and balanced enforcement strategies can promote responsible AI use \cite{lin2024beyond}.

Gen-AI is a transformative technology that enhances efficiency across disciplines, and fields that fail to adapt, risk falling behind. We recommend lenient use of Gen-AI in conferences, especially to support non-native English speakers. Gen-AI should enhance, not alter, the author’s work~\cite{lin2024beyond}, while reviewers' judgments must be independent. Both authors and reviewers should disclose Gen-AI use transparently and take full responsibility. Ultimately, scholars must verify content for accuracy and ethical compliance, as 
Gen-AI tools are imperfect and prone to hallucinations~\cite{naher-colm}.

\bibliographystyle{ACM-Reference-Format}
\bibliography{bibliography}

\end{document}